\documentclass[journal]{IEEEtran}

\ifCLASSINFOpdf
\usepackage[pdftex]{graphicx}

\else
\usepackage[dvips]{graphicx}
\graphicspath{{../eps/}}

\fi

\usepackage[cmex10]{amsmath}
\usepackage{caption}

\usepackage[table]{xcolor}
\usepackage{multirow}
\usepackage{multicol}
\usepackage{cuted}

\usepackage{enumerate}
\usepackage{subfigure}
\usepackage{acro}
\usepackage{epstopdf}
\usepackage{cite}
\usepackage{xcolor}

\usepackage{gensymb}

\usepackage{amsfonts}
\usepackage{siunitx}

\usepackage{caption}
\acsetup{list-long-format=\capitalisewords}

\DeclareAcronym{AOA}{short= AOA, long= angle of arrival}
\DeclareAcronym{AOD}{short= AOD, long= angle of departure}
\DeclareAcronym{CDF}{short= CDF, long= cumulative distribution function}
\DeclareAcronym{CSI}{short= CSI, long= channel state information}
\DeclareAcronym{CoMP}{short= CoMP, long= coordinated multipoint}
\DeclareAcronym{DL}{short= DL, long= downlink}
\DeclareAcronym{EM}{short= EM, long=  electromagnetic}
\DeclareAcronym{FCC}{short= FCC, long= Federal Communications Commission}
\DeclareAcronym{FoV}{short= FoV, long= field of view}
\DeclareAcronym{HO}{short= HO, long= handover}
\DeclareAcronym{LOS}{short= LOS, long= line-of-sight}
\DeclareAcronym{LTE}{short= LTE, long= long-term evolution}
\DeclareAcronym{mmwave}{short= mmWave, long= millimeter-wave}
\DeclareAcronym{MPC}{short= MPC, long= multipath component}
\DeclareAcronym{NLOS}{short= NLOS, long= non-line-of-sight}
\DeclareAcronym{OFDM}{short= OFDM, long= orthogonal frequency division multiplexing}
\DeclareAcronym{OFDMA}{short= OFDMA, long= Orthogonal Frequency Division Multiple Access}
\DeclareAcronym{PDF}{short= PDF, long=  probability density function}
\DeclareAcronym{PDP}{short= PDP, long=  power delay profile}
\DeclareAcronym{PSD}{short= PSD, long=  power spectral density}
\DeclareAcronym{RF}{short= RF, long=  radio frequency}
\DeclareAcronym{RMS}{short= RMS, long= root mean square}
\DeclareAcronym{RSRP}{short= RSRP, long= reference signal received power}
\DeclareAcronym{RVC}{short= RVC, long= Reverberation chamber}
\DeclareAcronym{SAW}{short= SAW, long=  surface acoustic wave}
\DeclareAcronym{SINR}{short= SINR, long= signal to interference plus noise ratio}
\DeclareAcronym{SON}{short= SON, long= self optimizing network}
\DeclareAcronym{TP}{short= TP, long= transmission point}
\DeclareAcronym{UL}{short= UL, long= uplink}
\DeclareAcronym{UE}{short= UE, long= user equipment}
\DeclareAcronym{IDT}{short= IDT, long= interdigital transducer}
\DeclareAcronym{CIR}{short= CIR, long= channel impulse response}
\DeclareAcronym{$LCR_f$}{short= $LCR_f$, long= level crossing rate}
\DeclareAcronym{ABF}{short= ABF, long= average bandwidth of fade}
\DeclareAcronym{INI}{short= INI, long= inter numerology interference}
\DeclareAcronym{SIR}{short= SIR, long= signal-to-interference ratio}
\DeclareAcronym{$P_{off}$}{short= $P_{off}$, long= power offset}
\DeclareAcronym{SCs}{short= SCs, long= subcarrier spacing}
\DeclareAcronym{LCM}{short= LCM, long= least common multiplier}
\DeclareAcronym{MSE-OFDM}{short= MSE-OFDM, long= multi symbols encapsulated OFDM}
\DeclareAcronym{FFT}{short= FFT, long= fast Fourier transform}
\DeclareAcronym{IFFT}{short= IFFT, long= inverse fast Fourier transform}
\DeclareAcronym{EVM}{short= EVM, long= error vector magnitude}
\DeclareAcronym{SSO}{short= SSO, long= subcarrier spacing offset}
\DeclareAcronym{PAPR}{short= PAPR, long= peak-to-average power ratio}
\DeclareAcronym{3GPP}{short= 3GPP, long= 3rd Generation Partnership Project}
\DeclareAcronym{ITU}{short= ITU-R, long= International Telecommunication Union Radiocommunication}
\DeclareAcronym{5G}{short= 5G, long= fifth generation}
\DeclareAcronym{URLLC}{short= URLLC, long= ultra-reliable low latency communication}
\DeclareAcronym{5G-NR}{short= 5G-NR, long= 5G New Radio}
\DeclareAcronym{eMBB}{short= eMBB, long= enhanced mobile broadband}
\DeclareAcronym{mMTC}{short= mMTC, long= massive machine type communication}
\DeclareAcronym{CP}{short= CP, long= cyclic prefix}
\DeclareAcronym{FBMC}{short= FBMC, long= filter bank multi-carrier}
\DeclareAcronym{GFDM}{short= GFDM, long= generalized frequency division multiplexing}
\DeclareAcronym{UFMC}{short= UFMC, long= universal filtered multi-carrier}
\DeclareAcronym{SFMC}{short= SFMC, long= subband filtered multi-carrier}
\DeclareAcronym{SM}{short= SM, long= subcarrier mapping}
\DeclareAcronym{NOMA}{short= NOMA, long= non-orthogonal multiple access}
\DeclareAcronym{ICI}{short= ICI, long= inter carrier interference}
\DeclareAcronym{ISI}{short= ISI, long= inter symbol interference}
\DeclareAcronym{BPSK}{short= BPSK, long= binary phase shift keying}

\begin{document}
	%
	\title{Multi-Numerology Multiplexing and Inter-Numerology Interference Analysis for 5G}

	\author{Abuu~B.~Kihero,
		Muhammad~Sohaib~J.~Solaija,~\IEEEmembership{Student Member,~IEEE}, and~H{\"u}seyin~Arslan,~\IEEEmembership{Fellow,~IEEE}
		
		\thanks{Authors are with School of Engineering and Natural Science, Istanbul Medipol University, Beykoz, 34810 Istanbul, TURKEY. H. Arslan is also with Department of Electrical Engineering, University of South Florida, Tampa, 33620 Florida, USA (e-mail: abkihero@st.medipol.edu.tr; msolaija@st.medipol.edu.tr; huseyinarslan@medipol.edu.tr).}
	}
	
	\maketitle
	
	\begin{abstract}
		Fifth generation (5G) radio access technology (RAT) is expected to flexibly serve multiple services with extremely diverse requirements. One of the steps taken toward fulfilling this vision of the 5G-RAT is an introduction of the multi-numerology concept, where multiple frame structures with different subcarrier spacings coexist in one frequency band. Though efficient in providing the required flexibility, this approach introduces a new kind of interference into the system known as inter numerology interference (INI). This study is geared toward analyzing the INI problem considering a cyclic prefix orthogonal frequency domain multiplexing (CP-OFDM) system and exposing the factors contributing to it through mathematical analyses. In-depth discussion of various critical issues concerning multi-numerology system such as frequency domain multiplexing and time domain symbol alignment for mixed numerologies is presented. Based on the findings of the conducted analyses, the paper highlights some approaches for minimizing INI in the system. The developed mathematical analysis is finally verified by Monte-Carlo simulations.
	\end{abstract}
	
	
	\IEEEpeerreviewmaketitle
\section{Introduction}
The vision for \ac{5G} of wireless technology is much more than the mere evolution of broadband services. It envisages a more diverse network with seamless coverage, higher data rates, various services, massive connectivity, and significantly higher reliability than any earlier generation of mobile communication. As compared to the \ac{LTE}, \ac{5G} advertises a thousand-fold increase in data volume and number of connected devices per area, 100 times improvement in user data rates, reduction of energy consumption by a factor of 10 and decrease in latency by a factor of 5 \cite{5g20155g}. These demands, when looked at simultaneously, seem improbable if not impossible to achieve, requiring multiple
magnitudes of improvement in current technology and infrastructure. 

As an approach to address these diverse service requirements, the standardization bodies, i.e., \ac{3GPP} and \ac{ITU}, have divided them into three different sets, one for each service, namely, \ac{eMBB}, \ac{URLLC}, and \ac{mMTC}. \ac{eMBB} prioritizes large bandwidth and high data rate, \ac{URLLC} needs high reliability and low latency while \ac{mMTC} necessitates low energy and bandwidth consumption and high coverage density\cite{zaidi2016waveform}. It is important to keep in mind that this division of services is by no means thorough, and there are going to be various scenarios that do not fall completely within a single defined use-cases\cite{dahlman20185g}. 

In order to meet these diverse requirements of \ac{5G} services, there have been two primary candidate approaches proposed by researchers. First, proposition of new waveforms for \ac{5G-NR} and second, using different numerologies of the same parent waveform, \ac{OFDM}, where the term \textit{numerology} refers to the different configurations of subcarrier spacing and \ac{CP} duration of an \ac{OFDM} symbol\cite{zhang2018mixed}. The first approach led to the proposition, analysis and comparison of various waveforms like \ac{FBMC}, \ac{GFDM}, and \ac{UFMC}. Several studies have compared these waveforms \cite{zhang2016waveform},\cite{gerzaguet20175g} and it is the general consensus among the researchers that while each of these waveforms offers improvement over \ac{OFDM} in some aspect, none of them can claim to address all requirements of \ac{5G}. In addition to this, \ac{OFDM}'s maturity and the immense effort that is already put into its standardization for \ac{LTE} led to it being selected as the waveform for \ac{5G} despite its \ac{CP} overhead, frequency and time offset sensitivity, high \ac{PAPR} and spectral regrowth issues\cite{5G_candidates_RS}. Therefore, for \ac{5G-NR}, \ac{3GPP} has opted for the second approach i.e. use of multiple numerologies of \ac{OFDM} waveform. Table \ref{tab:numerology_structure} gives a brief summary of the different numerologies present in the standard\cite{3GPP_38_211}.  A total of five scalable numerology options are provided with the conventional \ac{LTE} numerology (with subcarrier spacing of 15 kHz) chosen as the fundamental numerology. By scalable we mean that all the standardized $\Delta f$'s are $2^{\mu}$ multiples of the fundamental \ac{LTE} numerology (for $\mu = \{0,4\}$). One basic reason for choosing this set of numerologies with scalable subcarrier spacing is to facilitate easy symbols alignment and clock synchronization in time domain and thus simplifies the implementation \cite{zaidi2016waveform,vihriala2016numerology}.
\begin{table}[]
\centering
\caption{Numerology structures for data channels in 5G\cite{3GPP_38_211}.}
\label{tab:numerology_structure}
\begin{tabular}{|l|c|c|c|c|c|}
\hline
\multicolumn{1}{|c|}{\multirow{2}{*}{\textbf{Parameters}}}                              & \multicolumn{5}{c|}{\textbf{Numerology options} ($\mu$)}   \\ \cline{2-6} 
\multicolumn{1}{|c|}{}                                                                  & \textbf{0} & \textbf{1} & \textbf{2}  & \textbf{3} & \textbf{4} \\ \hline
\begin{tabular}[c]{@{}l@{}}Subcarrier Spacing\\  $(\si{\kilo\hertz})$\end{tabular}      & 15         & 30         & 60          & 120        & 240        \\ \hline
\begin{tabular}[c]{@{}l@{}}OFDM Symbol \\ Duration  $(\si{\micro\second})$\end{tabular} & 66.67      & 33.33      & 16.67       & 8.33       & 4.17       \\ \hline
\begin{tabular}[c]{@{}l@{}}CP Duration \\ $(\si{\micro\second})$\end{tabular}           & 4.69       & 2.34       & 4.17$|$1.17 & 0.58       & 0.29       \\ \hline
\begin{tabular}[c]{@{}l@{}}Slot Duration \\ $(\si{\milli\second})$\end{tabular}         & 1          & 0.5        & 0.25        & 0.125      & 0.0625     \\ \hline
\end{tabular}
\end{table}

It can be seen intuitively how different numerologies can be used to meet the demands of each service class of \ac{5G}. For example, lower (in terms of subcarrier spacing) numerologies are more suitable for \ac{mMTC}, since they can support higher number of simultaneously connected devices within the same bandwidth and require lower power, intermediate numerologies are appropriate for \ac{eMBB} which requires both, high data rate and significant bandwidth, and higher numerologies are more suitable for delay-sensitive applications pertaining to the \ac{URLLC} service due to shorter symbol duration. It is important to note that the selection of numerologies is not only dependent on the service to be supported, other factors like the cell size, time variation of the channel, delay spread, carrier frequency etc, also need to be taken into consideration in picking a numerology\cite{sachs20185g}.

\begin{figure*}
	\centering
	\subfigure[]{\includegraphics[width=1\columnwidth]{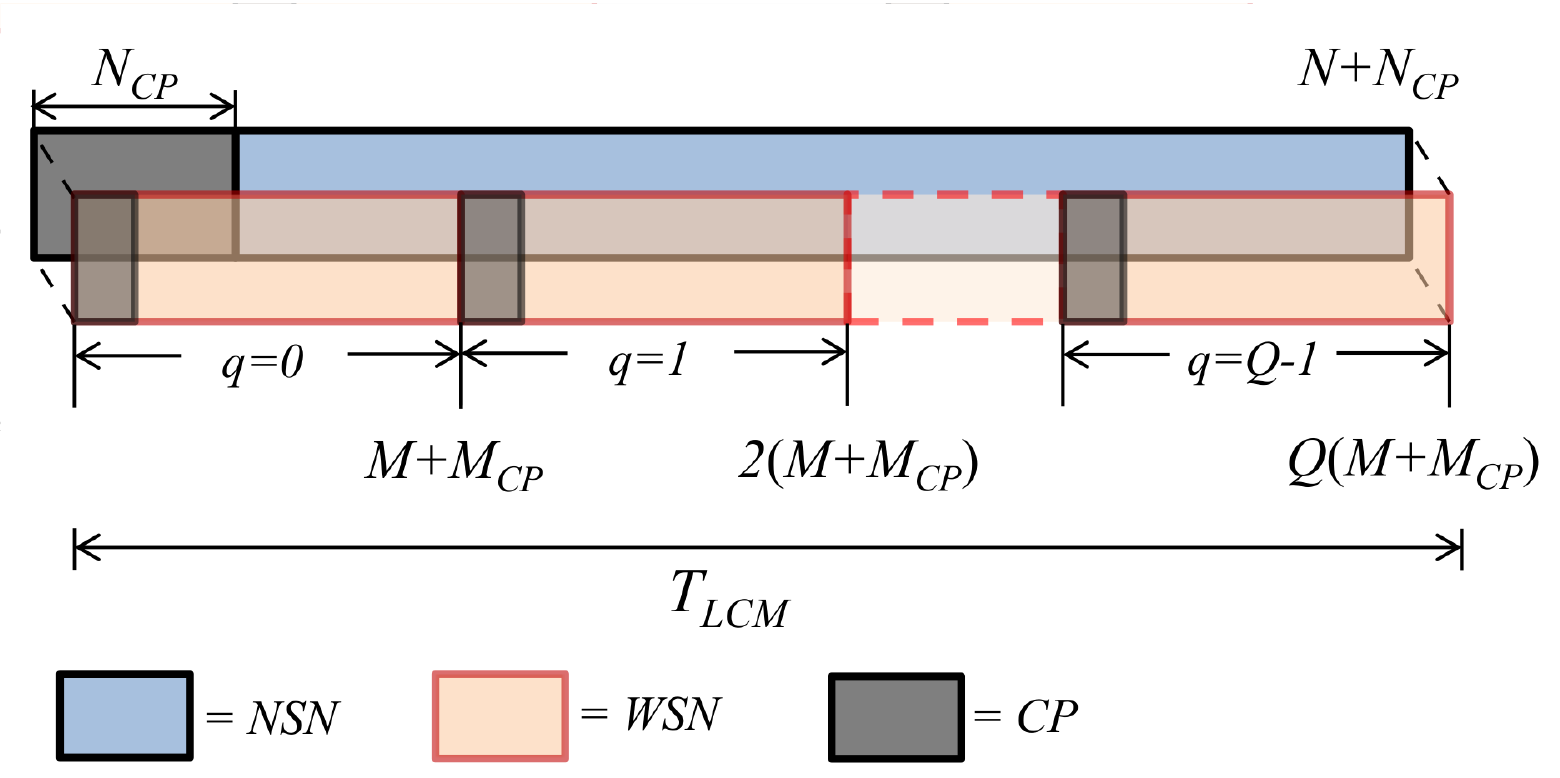} \label{fig:IndividualCP}
	}
	\subfigure[]{\includegraphics[width=1\columnwidth]{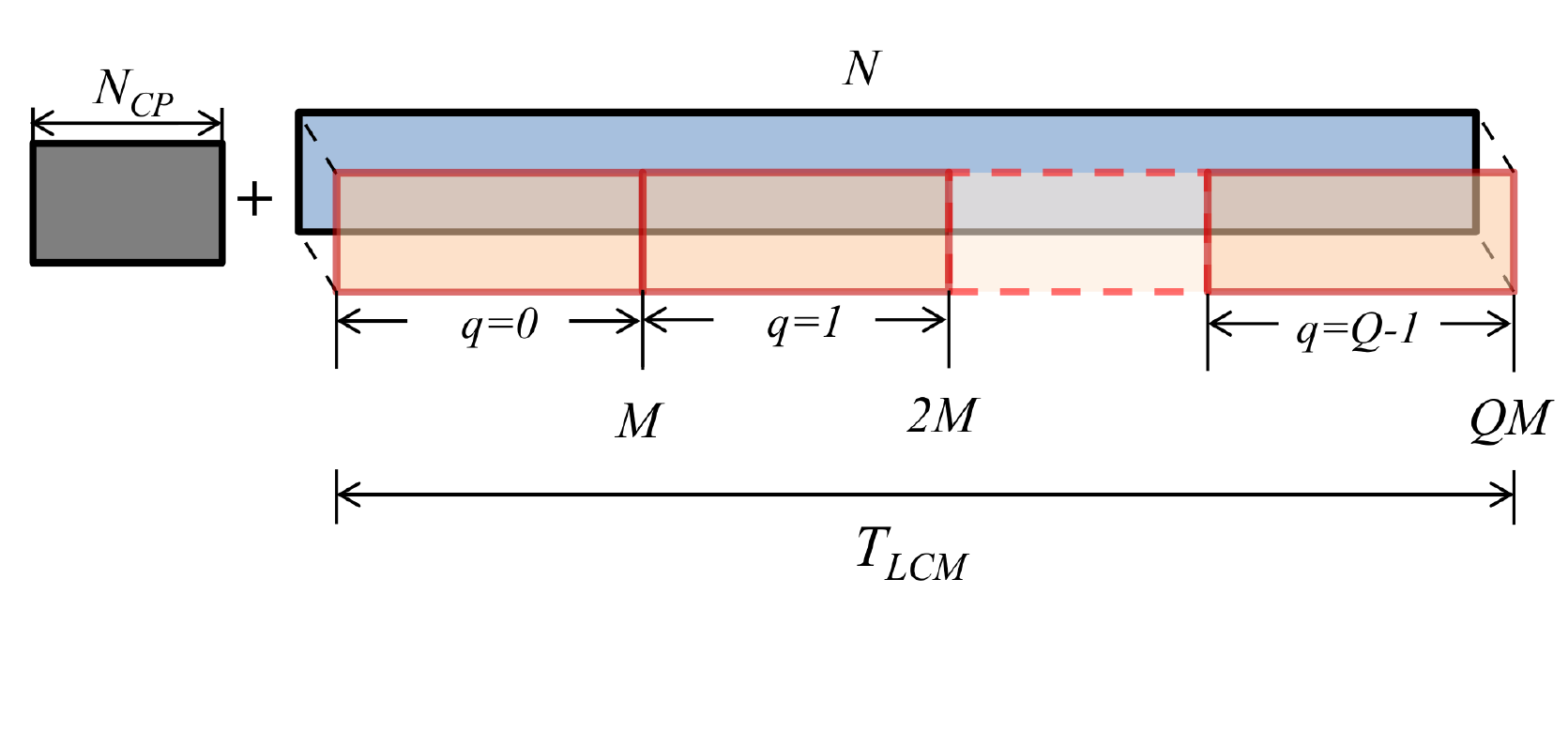} \label{fig:CommonCP}
	}
	\caption{Multi-numerology symbol alignment over $LCM$ symbol duration. (a) Individual CP case. (b) Common CP case. }
	\label{fig:SymbolAlignement}
\end{figure*}

While the use of multiple numerologies is essential in providing the necessary flexibility required for the diverse services, it introduces non-orthogonality into the system which in turns causes interference between the multiplexed numerologies \cite{zhang2017subband,zaidi2016waveform,zhang2018mixed,sachs20185g}. This new form of interference is referred to as \ac{INI}. Apart from causing loss of orthogonality among subcarriers in frequency domain, mixed numerologies also bring up difficulty in achieving symbols alignment in time domain. With the same sampling rate, an \ac{OFDM} symbol of one numerology does not perfectly align with the symbol of another numerology which makes synchronization issues within the frame difficult. However, with the scalable numerology design of 5G, symbol duration of one numerology is an integral multiple of the symbol duration of the other numerologies (see Fig. \ref{fig:SymbolAlignement}). Therefore, multi-numerology symbols can be perfectly aligned over the integral least common multiplier duration ($T_{LCM}$) as discussed in \cite{zhang2017subband,kihero2018inter,vihriala2016numerology,zhang2017multi}. Two different approaches of achieving symbol alignment over $T_{LCM}$ duration are presented in \cite{kihero2018inter}, namely individual CP and common CP. The individual CP refers to the conventional CP configuration where a CP is attached to each \ac{OFDM} symbol prior to transmission, as standardized for \ac{LTE} and 5G systems. Common CP, on the other hand, is a new CP configuration technique where one CP is used to protect multiple \ac{OFDM} symbols against \ac{ISI} \cite{wang2003comparison,chouinard2005mse,nemati2018low,kihero2018inter,abusabah2018noma}. It is a promising CP configuration technique for future wireless systems. It is briefly shown that these two approaches yield different INI distribution among subcarriers of the victim numerology. Proper understanding of the INI distribution is very crucial as it can open doors for developing better techniques of avoiding or minimizing INI in the system.

In order to make an intelligent use of the multiple numerologies concept introduced by \ac{5G-NR}, an in-depth analysis and understanding of INI is inevitable. Researchers have already taken a step toward describing \ac{INI} in details and identify various factors contributing to it.  A more generic study of the factors affecting INI is conducted in \cite{kihero2018inter}, but its scope is limited to the identification and intuitive explanation of the identified factors, supported by simulations without providing any analytical verification. A study in \cite{zhang2017subband} provides a more generic mathematical analysis of \ac{INI} for \ac{SFMC} based systems, and proposes some algorithms to minimize it. Zhang et. al. in \cite{zhang2018mixed} present a mathematical analysis of \ac{INI} and describe factors such as channel response, spectral distance between interfering and victim numerology, and windowing operation at the transmitter and receiver to have great impact on the amount of \ac{INI} suffered by the system. Though detailed, the focus of the presented analysis is limited to windowed-\ac{OFDM} system, and only an approximate analytical model of \ac{INI} for \ac{CP}-\ac{OFDM} is given. Apart from mere analysis of the \ac{INI}, some studies \cite{zhang2017subband,zhang2018mixed,demir2017impact} have already gone further and proposed techniques and algorithms to contain \ac{INI}. For instance, \cite{demir2017impact} tries to optimize the guard in time and frequency domains to minimize \ac{INI} while keeping in view the user requirements and their power offset. The study in \cite{Yazar2018Flexible} discuses a power offset based scheduling technique in multi-numerology systems to minimize \ac{INI} experienced by edge subcarriers of the multiplexed numerologies. 

Regarding the fact that the standardization bodies have selected CP-\ac{OFDM} as a standard waveform for \ac{5G-NR}, an extensive \ac{INI} analysis specific for this waveform is unarguably important in the literature. Our study, therefore, will focus on the \ac{INI} analysis for this core waveform of 5G. The analysis not only provides better understanding of the phenomenon, but also identifies areas where \ac{INI} can be either controlled/minimized or even exploited to enhance performance of the system. The contribution of this paper can be itemized as follows: 
\begin{itemize}
	\item A generic mathematical model for \ac{INI} in multi-numerology systems utilizing CP-\ac{OFDM} is provided, and analytical expressions for \ac{INI} are calculated. The established \ac{INI} models give clear insight on how different parameters contribute to \ac{INI} in the system. The developed \ac{INI} expressions are verified with Monte Carlo simulations. 
	\item A thorough investigation and comparison of the two symbols alignment techniques (i.e., individual and common CP) presented in \cite{kihero2018inter} are done from \ac{INI} perspective. The \ac{INI} mathematical analysis is done for both techniques.
\end{itemize}
Along with these main contributions, a generic method of multiplexing numerologies in a given spectrum band is described in detail from both frequency and time domains perspectives and their mathematical expressions are also given.

The rest of this paper is organized as follows. Section \ref{sec:system_model} discusses the downlink system model used for implementation of multi-numerology system and its mathematical formulation in both, frequency and time domains. Sections \ref{sec:INIanalysisIndividualCP} and \ref{sec:INIanalysisCommonCP} give detailed derivations of the INI expressions for individual and common CP configurations respectively. Numerical analysis and insights derived from the observed results are discussed in Section \ref{sec:discussion}. Section \ref{sec:conclusion} finally concludes the paper.

\section{System Model} \label{sec:system_model}
\begin{figure*}
    \centering
    \includegraphics[width=1.6\columnwidth]{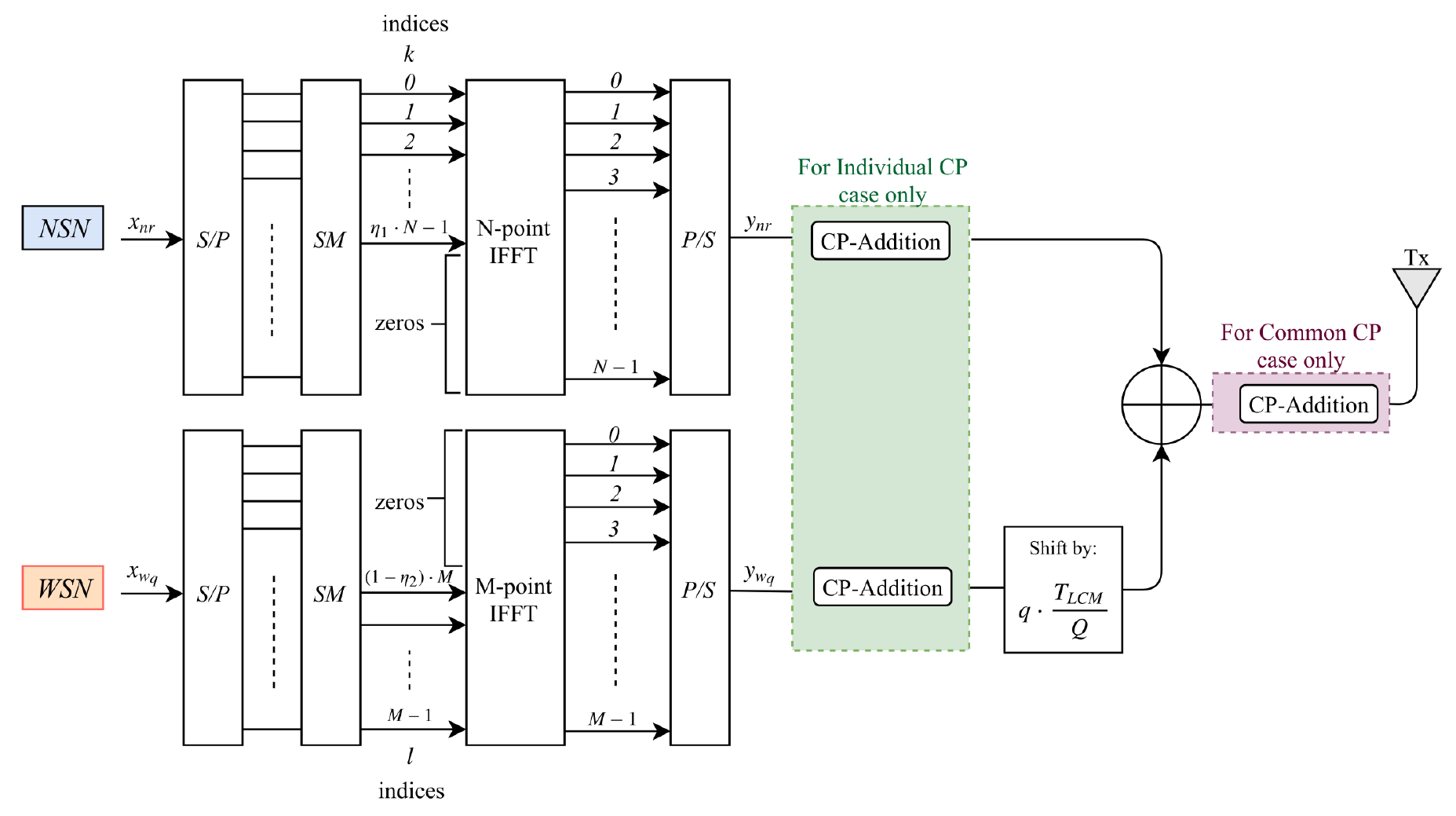}
    \caption{Downlink multi-numerology implementation at the transmitter.}
    \label{fig:multiplexedsubcarriers}
\end{figure*}

\subsection{Frequency Domain} \label{sec:SystemModelFreqDomain}
Let us consider a multi-numerology system with system bandwidth $B$ shared between two users. Active subcarriers of the two users are assumed to utilize the bandwidth in the ratio $\eta_1$ and $\eta_2$. For simplicity and numerical tractability of the INI analysis, only two users utilizing numerology-1 and numerology-2 are considered in this model. However, the analysis developed herein can be applied to any number of multiplexed numerologies by considering one pair of numerologies at a time. Let the two numerologies utilize subcarrier spacing $\Delta f_1$ (narrow subcarrier spacing) and $\Delta f_2$ (wide subcarrier spacing) respectively. Throughout the paper we will refer to the first numerology with narrow subcarrier spacing as NSN, and to the second with wide subcarrier spacing as WSN. Abiding by the 5G standards, the ratio $\Delta f_2/\Delta f_1$ = $Q$ is always an integer power of 2. Note that $Q=1$ refers to the case in which two users are using the same numerology.

Fig. \ref{fig:multiplexedsubcarriers} models the downlink implementation of the multi-numerology frequency-domain subcarriers multiplexing, time-domain symbol alignment, and creation of the composite signal for the transmission. The model employs \ac{SM} blocks right after serial-to-parallel (S/P) blocks for properly arranging NSN and WSN subcarriers according to their specified portions of the spectrum. The \ac{SM} blocks implement the localized subcarrier mapping technique whose output can be expressed as
\begin{equation}\label{eq:INIEq1}
X_{nr}(k) = \left\{
        \begin{array}{ll}
            x_{nr}(k) & \quad  ,0 \leq k \leq \eta_1N - 1\\
            0 & \quad ,\eta_1N \leq k \leq N - 1
        \end{array},
    \right.
\end{equation}
and
\begin{equation}\label{eq:INIEq2}
X_w(l) = \left\{
        \begin{array}{ll}
            0 & \quad ,0 \leq l \leq (1-\eta_2)M-1\\
            x_w(l) & \quad  ,(1-\eta_2)M \leq l \leq M - 1
        \end{array},
    \right.
\end{equation}
where, $N$ and $M$ are \ac{FFT}/\ac{IFFT} sizes of NSN and WSN respectively, such that $N = Q\cdot M$. $X_{nr}(k)$ and $X_w(l)$ are complex modulated symbols on $k^{th}$ and $l^{th}$ subcarriers of NSN and WSN respectively, after \ac{SM}. If we are to observe the WSN subcarriers with the granularity of the NSN subcarriers, the $l$-indices of active subcarriers of WSN can be defined as $l = \eta_1M + k/Q$ for $\{0 \leq k \leq \eta_2N - 1:k/Q \in \mathbb{Z}\}$.

\subsection{Time Domain} \label{sec:SystemModelTimeDomain}
The output of the \ac{IFFT} blocks in Fig. \ref{fig:multiplexedsubcarriers} are time-domain symbols $y_{nr}$ and $y_w$ of NSN and WSN respectively, and they are given by
\begin{equation}\label{eq:INIEq4}
\begin{split}
y_{nr}(n) &= IFFT\{X_{nr}\}|_{N-points} \\&
= \frac{1}{\sqrt{N}}\sum_{k=0}^{\eta_1 N-1}X_{nr}(k)\mathrm{e}^{j2\pi nk/N},\\&
\quad \quad for \quad 0 \leq n \leq N - 1,
\end{split}
\end{equation}
and
\begin{equation}\label{eq:INIEq5}
\begin{split}
y_w(m) &= IFFT\{X_{w}\}|_{M-points} \\&
= \frac{1}{\sqrt{M}}\sum_{k=0}^{\eta_2N-1}X_{w}(\eta_1M + k/Q)\mathrm{e}^{\frac{j2\pi}{N} m(k+\eta_1N)}\\&
\quad \quad for \quad 0 \leq m \leq M - 1, \quad k/Q \in \mathbb{Z}.
\end{split}
\end{equation}

As briefly explained in the introduction, the symbol alignment issue in the multi-numerology system is overcome when the multiplexed numerologies are integral multiple of one another. For instance, subcarrier spacings $\Delta f_2$ = $Q\cdot\Delta f_1$ imply that the symbol durations $T_1$ and $T_2$ are also related by $T_1 = Q\cdot T_2$, which means that $Q$-concatenated symbols of WSN can be perfectly aligned with one symbol of NSN. This gives rise to the concept of multi-numerology symbols alignment over $T_{LCM}$ duration as mentioned earlier. $T_{LCM}$ is usually equal to the symbol duration of the numerology with the smallest $\Delta f$ in a given system. With an exception of the extended CP option for the numerology with $\Delta f$ = 60 kHz, all standardized numerologies use the same CP-ratio, which guarantees that the multi-numerology symbols alignment over $T_{LCM}$ duration holds even after CP addition to the symbols. As we have mentioned before, in this study individual CP and common CP approaches of achieving multi-numerology symbols alignment are investigated.

\subsubsection{Individual CP}
The individual CP configuration is shown in Fig. \ref{fig:IndividualCP}. It is a conventional approach where CP is added to each of the $Q$ symbols of WSN before concatenating them. Let $N_{CP}$ and $M_{CP}$ be the CP-sizes for NSN and WSN respectively, where $N_{CP} = Q\cdot M_{CP}$. After CP addition to each symbol, the ranges of indices $n$ and $m$ in (\ref{eq:INIEq4}) and (\ref{eq:INIEq5}) respectively are redefined to $-N_{CP} \leq n \leq N-1$, and $-M_{CP} \leq m \leq M-1$. The $Q$-concatenated symbols of WSN (denoted by $y_{w_c}^I$) can be modeled as an $M$-point periodic extension of $y_w$, given as
\begin{equation}\label{eq:INIEq6}
\begin{split}
y_{w_c}^I = \sum_{q=0}^{Q-1}y_{w_q}&(m-q(M+M_{CP})),
\end{split}
\end{equation}
where $q$ is the WSN symbol index within $T_{LCM}$ duration. Now, $y_{nr}$ and $y_{w_c}^I$ are summed up to form a composite signal $y$ of the two numerologies for transmission \cite{zhang2017multi}, given as
\begin{equation}\label{eq:INIEq7}
\begin{split}
y = y_{nr} + y_{w_c}^I.
\end{split}
\end{equation}

\subsubsection{Common CP}
In this approach all $Q$ concatenated WSN symbols are protected by only one CP of length $Q\times M_{CP}$ (= $N_{CP}$) as shown in Fig. \ref{fig:CommonCP}. This approach is motivated by the concept of multi-symbols encapsulated \ac{OFDM} (MSE-OFDM) studied in \cite{chouinard2005mse,wang2003comparison,nemati2018low}. As reported in Table \ref{tab:numerology_structure}, \ac{CP} size decreases with an increase in the subcarrier spacing. In case of a highly doubly dispersive channel, we do not have flexibility to utilize larger subcarrier spacing and larger \ac{CP} duration simultaneously to overcome \ac{ICI} and \ac{ISI} respectively. Common CP configuration provide a solution to this dilemma by allowing us to use larger subcarriers spacing and yet have larger \ac{CP} duration to contain the \ac{ISI} without compromising the spectral efficiency and receiver complexity \cite{nemati2018low}. In another study, authors have used common CP configuration to enhance performance of the multi-numerology based \ac{NOMA} systems \cite{abusabah2018noma}. Although common CP configuration have exhibited better performance compared to the conventional individual CP structure in some scenarios, its \ac{INI} performance is not yet well investigated. It should be understood that, a modified frequency domain equalizer has to be used at the receiver when common CP is used. Details concerning equalization with common CP can be found in \cite{wang2003comparison} and \cite{nemati2018low}.

Now, with the common CP, the $Q$-concatenated symbols of WSN (denoted by $y_{w_c}^C$) is given as
\begin{equation}\label{eq:INIEq26}
\begin{split}
y_{w_c}^C = \sum_{q=0}^{Q-1}y_{w_q}(m-N_{CP}-qM),
\end{split}
\end{equation}
and the transmitted signal $y$ is given by
\begin{equation}\label{eq:INIEq77}
\begin{split}
y = y_{nr} + y_{w_c}^C.
\end{split}
\end{equation}

Let us assume that the transmitted signal passes through  a $D$-taps time varying multipath channel with its discrete channel impulse response (CIR) given by
\begin{equation}\label{eq:INIEq8}
  h(m) = \sum_{d=0}^{D-1}\gamma_d(m)\delta(m-\tau_d),
\end{equation}
where $\tau_d$ and $\gamma_d$ are propagation delays (in samples) and complex amplitude associated with $d^{th}$ tap, respectively. $\delta(\cdot)$ is the Dirac delta function that models the propagation delay of each tap. The received signal $s$ is then expressed as
\begin{equation}\label{eq:INIEq9}
\begin{split}
  s(r) = h(r)*y(r) = \sum_{d=0}^{D-1}\gamma_d(r)y(r-\tau_d),
\end{split}
\end{equation}
where additive channel noise is ignored.

In this analysis, we assume that a proper choice of numerology is made such that neither NSN nor WSN subcarriers are affected by \ac{ICI}. Furthermore, we consider that $N_{CP}>D$ and $M_{CP}>D$ such that both numerologies are not affected by \ac{ISI} \cite{Hwang2009}. Therefore, only \ac{INI} is expected in the system.

\section{Inter-Numerology Interference Analysis for Individual CP Configuration} \label{sec:INIanalysisIndividualCP}
\subsection{INI from WSN to NSN} \label{sec:WSNtoNSN}
At the receiver of the NSN user, $N_{CP}$-sized CP is removed and then the time-domain received signal is transformed to its frequency domain for the frequency domain channel equalization and data detection processes. The N-point FFT of the received signal is taken as follows
\begin{equation}\label{eq:INIEq10}
\begin{split}
  S(v) &= FFT\{s(r)\}|_{N-point} = H^{NSN}(v)\cdot Y(v),
 \end{split}
\end{equation}
for $0 \leq v \leq \eta_1 \cdot N - 1$, where $v$ represents indices of the desired subcarriers of NSN, and $H^{NSN}(v)$ denotes the channel frequency response (CFR) at $v^{th}$ subcarrier, defined as
\begin{equation}\label{eq:INIEq11}
  H^{NSN}(v) = \sum_{d=0}^{D-1}\gamma^{NSN}_d(v)\mathrm{e}^{-\frac{j2\pi}{N}v\tau_d}.
\end{equation}
$Y$ is the FFT of the composite transmitted signal $y$, given as 
\begin{equation}\label{eq:INIEq12}
\begin{split}
  Y(v) &= \frac{1}{\sqrt{N}}\sum_{r=N_{CP}}^{N+N_{CP}-1}y(r)\mathrm{e}^{-\frac{j2\pi}{N} rv}.
\end{split}
\end{equation}
Since $y$ is composed of the signals of the two multiplexed numerologies, $Y$ will then consist of the desired symbols ($Y_{des}$) of NSN as well as the interference $Y_{INI}$ they receive from sidelobes of the WSN subcarriers. So, \eqref{eq:INIEq10} becomes
\begin{equation}\label{eq:INIEq10b}
\begin{split}
 S(v) = H^{NSN}(v)\cdot [Y_{des}(v) + Y_{INI}(v)].
 \end{split}
\end{equation}
Since we have considered the flat fading subchannel condition each subcarrier, then, the channel effect in \eqref{eq:INIEq10b} can be completely compensated by using a conventional one-tap frequency domain equalization.

By using \eqref{eq:INIEq12}, \eqref{eq:INIEq7}, and \eqref{eq:INIEq6}, we can express the interference $Y_{INI}(v)$ from WSN to $v^{th}$ subcarrier of NSN as
\begin{equation}\label{eq:INIEq15}
\begin{split}
Y_{INI}(v) &=\frac{1}{\sqrt{N}}\sum_{r=N_{CP}}^{M+M_{CP}-1}\sum_{q=0}^{Q-1}y_{w_q}(r-q(M+M_{CP}))\\&
\quad \quad
\cdot\mathrm{e}^{-\frac{j2\pi}{N}v(r-q(M+M_{CP}))}.
\end{split}
\end{equation}
Note that, among the $Q$ concatenated symbols of WSN, only the first symbol (i.e., $q=0$) overlaps with the CP portion of the NSN symbol. Therefore, due to the CP removal at the NSN receiver, we can write
\begin{equation}\label{eq:INIEq16}
\begin{split}
 &Y_{INI}(v) = \frac{1}{\sqrt{N}}\Bigg[\sum_{r=N_{CP}}^{M+M_{CP}-1}y_{w_0}(r)\mathrm{e}^{-\frac{j2\pi}{N}rv} + \\&
 \sum_{r=0}^{M+M_{CP}-1}\sum_{q=1}^{Q-1}y_{w_q}(r-q(M+M_{CP}))\mathrm{e}^{-\frac{j2\pi}{N}v (r-q(M+M_{CP}))}\Bigg]
\end{split}
\end{equation}
Further simplification of \eqref{eq:INIEq16} is shown in Appendix \ref{sec:AppendixA}. Now, by substituting \eqref{eq:INIEqAppA1} and \eqref{eq:INIEqAppA2} from the Appendix \ref{sec:AppendixA} into \eqref{eq:INIEq16}, we find the INI power $I_{NSN}(k,v)$ from $k^{th}$ subcarrier of WSN to the $v^{th}$ subcarrier of NSN as

\begin{equation}\label{eq:INIEq17}
\begin{split}
&I_{NSN}(k,v) = |Y_{INI_k}(v)|^2 = \frac{\rho^{WSN}(k)}{N\cdot M}\Psi(k,v),\\&
for \quad \quad 0\leq v \leq \eta_1N -1 \quad and \\& \quad \quad \quad \quad \{0\leq k \leq \eta_2N -1 : k/Q \in \mathbb{Z}\},
\end{split}
\end{equation}
where
\begin{equation}\label{eq:INIEq17b}
\nonumber
\begin{split}
\Psi(k,v) =&\frac{\Bigg|sin\bigg[\frac{\pi}{Q}\bigg(1+(1-Q)CP_R\bigg)(k-v) \bigg]\Bigg|^2}{\Bigg|sin\bigg[\frac{\pi}{N}(k-v+\eta_1N)\bigg]\Bigg|^2}  \\& +(Q-1)\cdot \frac{\Bigg|sin\bigg[\frac{\pi}{Q}(1+CP_R)(k-v) \bigg]\Bigg|^2}{\Bigg|sin\bigg[\frac{\pi}{N}(k-v+\eta_1N)\bigg]\Bigg|^2},
\end{split}
\end{equation}
$\rho^{WSN}(k)$ = $|X_{w}(\eta_2 M + k/Q)|^2$ is the WSN subcarrier power, and $CP_R$ = $N_{CP}/N$ = $M_{CP}/M$ is the CP ratio employed by the the system. The term $k-v+\eta_1N$ is the spectral distance between the interfering subcarrier at $k+\eta_1N$ and the victim subcarrier at $v$. Note that when $Q=1$, i.e., the system is comprised of a single numerology, the INI power in \eqref{eq:INIEq17} becomes zero as it would be expected. \Ac{SIR} performance of the $v^{th}$ subcarrier of NSN due to $I_{NSN}(k,v)$ can be calculated as
\begin{equation}\label{eq:SIReq1}
\begin{split}
SIR_{NSN}(v) &= \rho^{NSN}(v)\bigg/\sum_{k=0}^{\eta_2N -1}I_{NSN}(k,v)\\&
= \frac{NM\cdot\rho^{NSN}(v)}{\sum_{k=0}^{\eta_2N -1}\rho^{WSN}(k) \Psi(k,v)}.
\end{split}
\end{equation}

\subsection{INI from NSN to WSN} \label{sec:NSNtoWSN}
Again, we consider $s(r)$ in \eqref{eq:INIEq9} as the received signal at WSN receiver. To demodulate a $q^{th}$ symbol out of the $Q$-concatenated symbols of WSN, the WSN receiver captures an $s(r-q(M+M_{CP}))$ portion of the received signal, where $0\leq r \leq M+M_{CP}-1$, such that
\begin{equation}\label{eq:INIEq18}
 s(r-q(M+M_{CP})) = y_{w_q}(r) + y_{nr}(r-q(M+M_{CP})),
\end{equation}
where $y_{nr}(r-q(M+M_{CP}))$ is the portion of the NSN symbol added to the $q^{th}$ symbol of WSN during creation of the composite signal at the transmitter. M-point FFT process at the WSN receiver results in
\begin{equation}\label{eq:INIEq19}
\begin{split}
S(v) &= FFT\{s(r-q(M+M_{CP}))\}|_{M-point}\\&
= H^{WSN}(v)[Y_{des}(v) + Y_{INI}(v)],
\end{split}
\end{equation}
where $H^{WSN}(v)$ is the CFR at the $v^{th}$ subcarrier of WSN user given as 
\begin{equation}\label{eq:INIEq20}
  H^{WSN}(v) = \sum_{d=0}^{D-1}\gamma^{WSN}_d(v)\mathrm{e}^{-\frac{j2\pi v\tau_d}{M}}.
\end{equation}
Recalling the localized subcarrier mapping technique implemented at the transmitter, the indices, $v$ of the WSN subcarriers to be detected can be expressed as $v = \eta_2M + p/Q$, where $\{0\leq p \leq \eta_2N -1 : p/Q \in \mathbb{Z}\}$. We can thus write \eqref{eq:INIEq19} as
\begin{equation}\label{eq:INIEq21}
\begin{split}
S(\eta_1M + p/Q) =& H^{WSN}(\eta_1M + p/Q)\bigg[Y_{des}(\eta_1M + p/Q) \\&+ Y_{INI}(\eta_1M + p/Q)\bigg].
\end{split}
\end{equation}
Again, considering perfect channel equalization at the receiver, we can get rid of the channel effect $H^{WSN}$ in \eqref{eq:INIEq21}.

Now, we can find the interference $Y_{INI}$ from NSN to WSN as
\begin{equation}\label{eq:INIEq24}
\begin{split}
&Y_{INI}(\eta_1M + p/Q)\\& = \frac{1}{\sqrt{M}}\sum_{r=M_{CP}}^{M+M_{CP}-1}y_{nr}(r-q(M+M_{CP}))\mathrm{e}^{-\frac{j2\pi}{M} r(\eta_1M + p/Q)}\\&
= \frac{1}{\sqrt{M}}\sum_{r=0}^{M-1}y_{nr}(r+M_{CP}-q(M+M_{CP}))\\& \quad \quad \quad \quad \quad \quad \quad 
\quad \quad \quad \quad 
\cdot\mathrm{e}^{-\frac{j2\pi}{M}(r+M_{CP})(\eta_1M + p/Q)}
\end{split}
\end{equation}
\eqref{eq:INIEq24} is simplified in Appendix \ref{sec:AppendixB} and \eqref{eq:INIEqAppA3} is obtained. Now, by using \eqref{eq:INIEq21} and \eqref{eq:INIEqAppA3} the INI power, $I_{WSN}(k,p)$ from $k^{th}$ subcarrier of NSN to the $p^{th}$ subcarrier of WSN can be given as
\begin{equation}\label{eq:INIEq25}
\begin{split}
I_{WSN}(k,p) &= |Y_{INI_k}(\eta_1M + p/Q)|^2 = \frac{\rho^{NSN}(k)}{N\cdot M}|\xi(k,p)|^2,\\&
for \quad 0\leq k \leq \eta_1N -1 \quad and \\& \quad \quad \{0\leq p \leq \eta_2N -1 : p/Q \in \mathbb{Z}\},
\end{split}
\end{equation}
where 
\begin{equation}
\nonumber
\begin{split}
\xi(k,p) = \frac{sin \bigg[\frac{\pi}{Q}(k-p)\bigg]}{sin \bigg[\frac{\pi}{N}(k-p-\eta_1N)\bigg]},
\end{split}
\end{equation}
and $\rho^{NSN}(k)$ = $|X_{nr}(k)|^2$ is the NSN subcarrier power. Again, note that, when only one numerology occupies the whole spectrum (i.e., $Q=1$), the $I_{WSN}$ is zero as expected. 

The SIR of $p^{th}$ subcarrier of WSN due to $I_{WSN}(k,p)$ is given by
\begin{equation}\label{eq:SIReq2}
\begin{split}
SIR_{WSN}(p) &= \rho^{WSN}(p)\bigg/\sum_{k=0}^{\eta_1N -1}I_{WSN}(k,p)\\&
= \frac{NM\cdot\rho^{WSN}(p)}{\sum_{k=0}^{\eta_1N -1}\rho^{NSN}(k)|\xi(k,p)|^2}.
\end{split}
\end{equation}

\textit{Remark 1:} From the INI expressions obtained in \eqref{eq:INIEq17} and \eqref{eq:INIEq25}, we can deduce that power $\rho$ of the interfering numerology, spectrum sharing factor $\eta $, as well as the ratio $\Delta f_2/\Delta f_1 = N/M = Q$ of the multiplexed numerologies are the main factors affecting the amount of INI experienced by each numerology. One can easily note that, if we increase $Q$ (i.e., decreasing M for a particular value of N) we minimize the product $N\cdot M$ in \eqref{eq:INIEq17} and \eqref{eq:INIEq25} resulting in a higher INI for both numerologies. This property was also observed by Zhang in \cite{zhang2018mixed}. This suggests that, for a given set of numerologies, system INI can be minimized by scheduling numerologies with minimum $Q$ adjacent to each other. 

\textit{Remark 2:} The SIR expressions \eqref{eq:SIReq1} and \eqref{eq:SIReq2} show that any power offset between the two numerologies favor SIR performance of one numerology (with relatively higher power) and deteriorate the other (with low power). The fairer case is observed when the multiplexed numerologies have the same power (i.e., $\rho^{NSN} = \rho^{WSN}$), where SIR performance of both numerologies becomes independent of their powers. These analytical relationships provide the basis of the heuristic multi-numerology scheduling algorithms proposed in \cite{demir2017impact} and \cite{Yazar2018Flexible} where the authors suggest that numerologies with minimum power offsets should be scheduled adjacent to each other. 

\section{Inter-Numerology Interference Analysis for Common CP Configuration}\label{sec:INIanalysisCommonCP}
While following the same derivation steps as detailed in Section \ref{sec:INIanalysisIndividualCP}, here in this section we will only summarize the INI derivations for WSN and NSN.

\subsection{INI from WSN to NSN} \label{sec:CommonCP-WSNtoNSN}
The $N$-points FFT processing at the NSN receiver yields the interference $Y_{INI}(v)$ from WSN to $v^{th}$ subcarrier of NSN as
\begin{equation}\label{eq:INIEq28}
\begin{split}
	Y_{INI}(v) &= \frac{1}{\sqrt{N}}\sum_{r=N_{CP}}^{N+N_{CP}-1}y_{w_c}^C(r)\mathrm{e}^{-\frac{j2\pi}{N}rv}\\&
	=\frac{1}{\sqrt{N}}\sum_{r=0}^{N-1}y_{w_c}^C(r+N_{CP})\mathrm{e}^{-\frac{j2\pi}{N}(r+N_{CP})v}.
\end{split}
\end{equation}
By using \eqref{eq:INIEq26} and \eqref{eq:INIEq5} and adopting the same step as in Appendix \ref{sec:AppendixA} and \ref{sec:AppendixB}, \eqref{eq:INIEq28} can be simplified to
\begin{equation}\label{eq:INIEq29}
\begin{split}
	Y_{INI}(v) =&\frac{1}{\sqrt{N\cdot M}}\sum_{q=0}^{Q-1}\sum_{k=0}^{\eta2N-1}X_{w_q}(\eta_1M + k/Q)\\& 
	\cdot\mathrm{e}^{-\frac{j2\pi q}{Q}r(k-v+\eta_1N}\mathrm{e}^{-\frac{j\pi}{N}(M-1)(k-v+\eta_1N)}
	\cdot \zeta(k,v)
\end{split}
\end{equation}
where
\begin{equation}
\nonumber
\begin{split}
\zeta(k,p) = \frac{sin \bigg[\frac{\pi}{Q}(k-v)\bigg]}{sin \bigg[\frac{\pi}{N}(k-v+\eta_1N)\bigg]}.
\end{split}
\end{equation}
Now, the INI power $I_{NSN}(k,v)$ from $k^{th}$ subcarrier of WSN to $v^{th}$ subcarrier of NSN is given as
\begin{equation}\label{eq:INIEq30}
\begin{split}
 I_{NSN}(k,v) &= |Y_{INI_k}(v)|^2 = \frac{\rho^{WSN}(k)}{M^2}|\zeta(k,v)|^2,\\&
for \quad 0\leq v \leq \eta_1N -1 \quad and \\& \quad \quad \quad \quad \{0\leq k \leq \eta_2N -1 : k/Q \in \mathbb{Z}\}.
\end{split}
\end{equation}
The SIR of $v^{th}$ subcarrier of NSN due to $I_{NSN}(k,v)$ is given as
\begin{equation}\label{eq:SIReq3}
\begin{split}
SIR_{NSN}(v) &= \rho^{NSN}\bigg/\sum_{k=0}^{\eta_2N-1}I_{NSN}(k,v)\\&
= \frac{M^2\rho^{NSN}(v)}{\sum_{k=0}^{\eta_2N-1}\rho^{WSN}(k)|\zeta(k,v)|^2}.
\end{split}
\end{equation}

\subsection{INI from NSN to WSN} \label{sec:CommonCP-NSNtoWSN}
Contrary to the previous case of WSN with individual CP where equalization is done for each of the $Q$-concatenated symbols, in this case, since all symbols share one CP, equalization is therefore done over the whole block of $Q$ WSN symbols. The equalized frequency domain signal is then converted back into the time domain where each symbol is demodulated through M-points FFT process. As we have mentioned before, further details concerning equalization with common CP can be found in \cite{wang2003comparison} and \cite{nemati2018low}. Again, by considering perfect equalization, we calculate the interference $Y_{INI}$ from NSN to WSN as 
 \begin{equation}\label{eq:INIEq31}
\begin{split}
	Y_{INI}&(\eta_1M + p/Q) \\&
	=\frac{1}{\sqrt{M}}\sum_{r=0}^{M-1}y_{nr}(r-qM)\mathrm{e}^{-\frac{j2\pi}{N}(r-qM)(\eta_1N + p)}.
\end{split}
\end{equation}
Substituting \eqref{eq:INIEq4} into \eqref{eq:INIEq31} and perform some simplifications through similar steps used in Appendix \ref{sec:AppendixA} and \ref{sec:AppendixB}, an expression for the INI power, $I_{WSN}(k,p)$ from $k^{th}$ subcarrier of NSN to $p^{th}$ subcarrier of WSN can be found as
\begin{equation}\label{eq:INIEq32}
\begin{split}
I_{WSN}(k,p) &= |Y_{INI_k}(\eta_1M + p/Q)|^2\\&
=\frac{\rho^{NSN}(k)}{N\cdot M}|\xi(k,p)|^2,\\&
for \quad \quad 0\leq k \leq \eta_1N -1 \quad and \\& \quad \quad \quad \quad \{0\leq p \leq \eta_2N -1 : p/Q \in \mathbb{Z}\}.
\end{split}
\end{equation}
One can notice from \eqref{eq:INIEq25} and \eqref{eq:INIEq32} that the INI power leaking from NSN to WSN for the individual and common CP cases are exactly the same. Therefore, their SIR performance are also the same, given by \eqref{eq:SIReq2}. In this regard, we can conclude that changing CP configuration affects INI distribution of the NSN only.

One difference between the INI characteristics of the two investigated CP configurations is that, the INI experienced by NSN for the individual CP is a function of $CP_R$, among other factors, while that is not the case with the common CP. This is intuitively plausible as in the common CP configuration no CP portion of WSN concatenated symbols overlaps with the symbol portion of the NSN (see Fig. \ref{fig:SymbolAlignement}).

\section{Numerical Results and Discussion} \label{sec:discussion}
In this section we use Monte-Carlo simulations to evaluate the accuracy of the analytically derived INI expressions. A multi-numerology system utilizing CP-OFDM waveform with NSN and WSN numerologies equally sharing the available bandwidth (i.e., $\eta_1 = \eta_2$ = 0.5) is considered. Throughout the analysis a \ac{CP} ratio of 1/16, N = 128, and \ac{BPSK} modulated symbols with unit power are used, unless otherwise specified. As we have observed from the analytic expression in the previous sections, the amount of INI experienced by each numerology in the system depends on the ratio $Q$ of the subcarrier spacings of the multiplexed numerologies instead of the actual values of their subcarrier spacing. Therefore, we also consider evaluating performance of the system for different values of $Q$. 
\begin{figure}
	\centering
	\includegraphics[width=1\columnwidth]{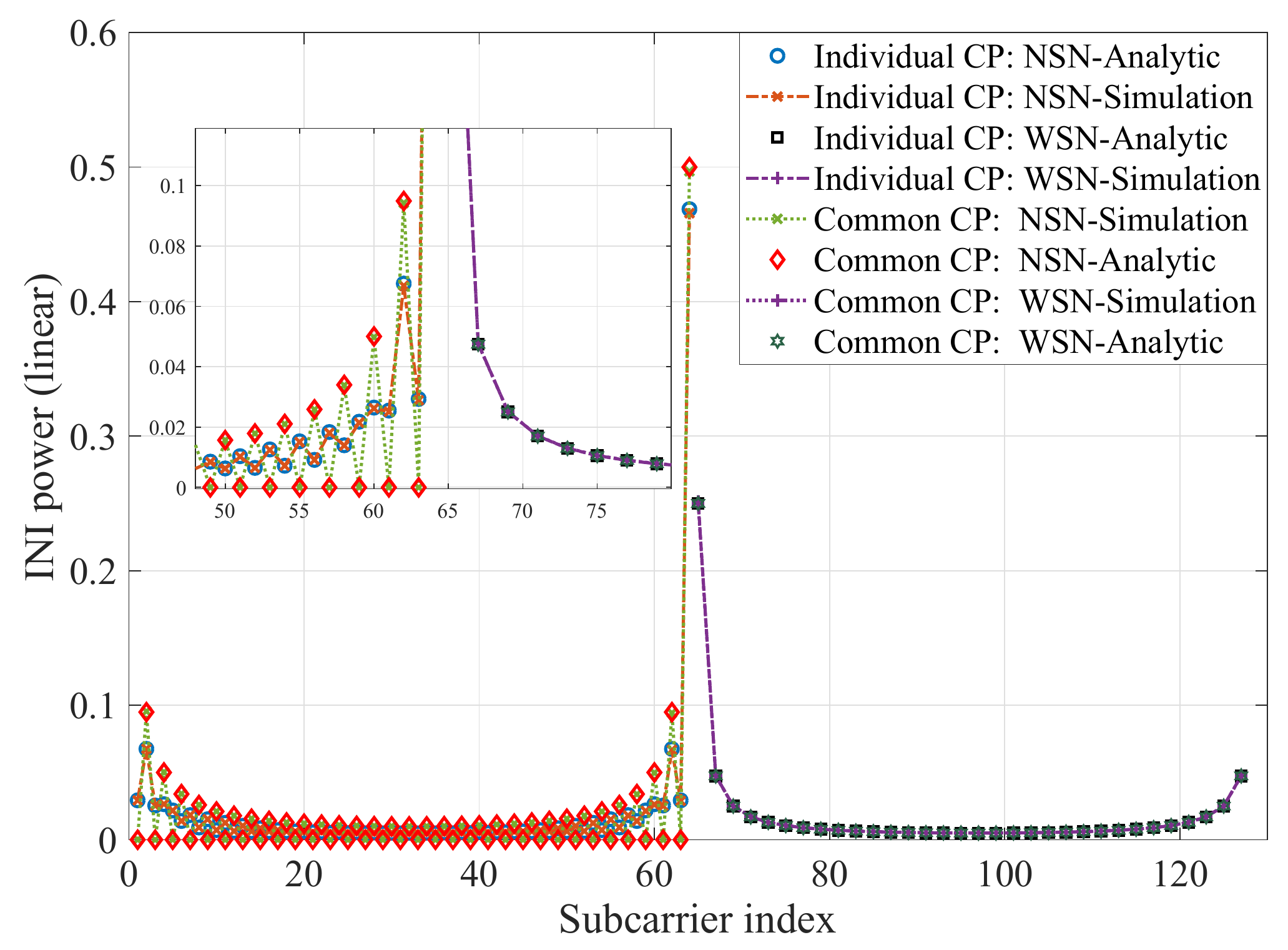}
	\caption{INI distribution on NSN and WSN subcarriers for individual and common CP configurations}
	\label{fig:IndividualCommonCP}
\end{figure}
\begin{figure*}
	\centering
	\subfigure[Individual CP]{
		\includegraphics[width=0.95\columnwidth]{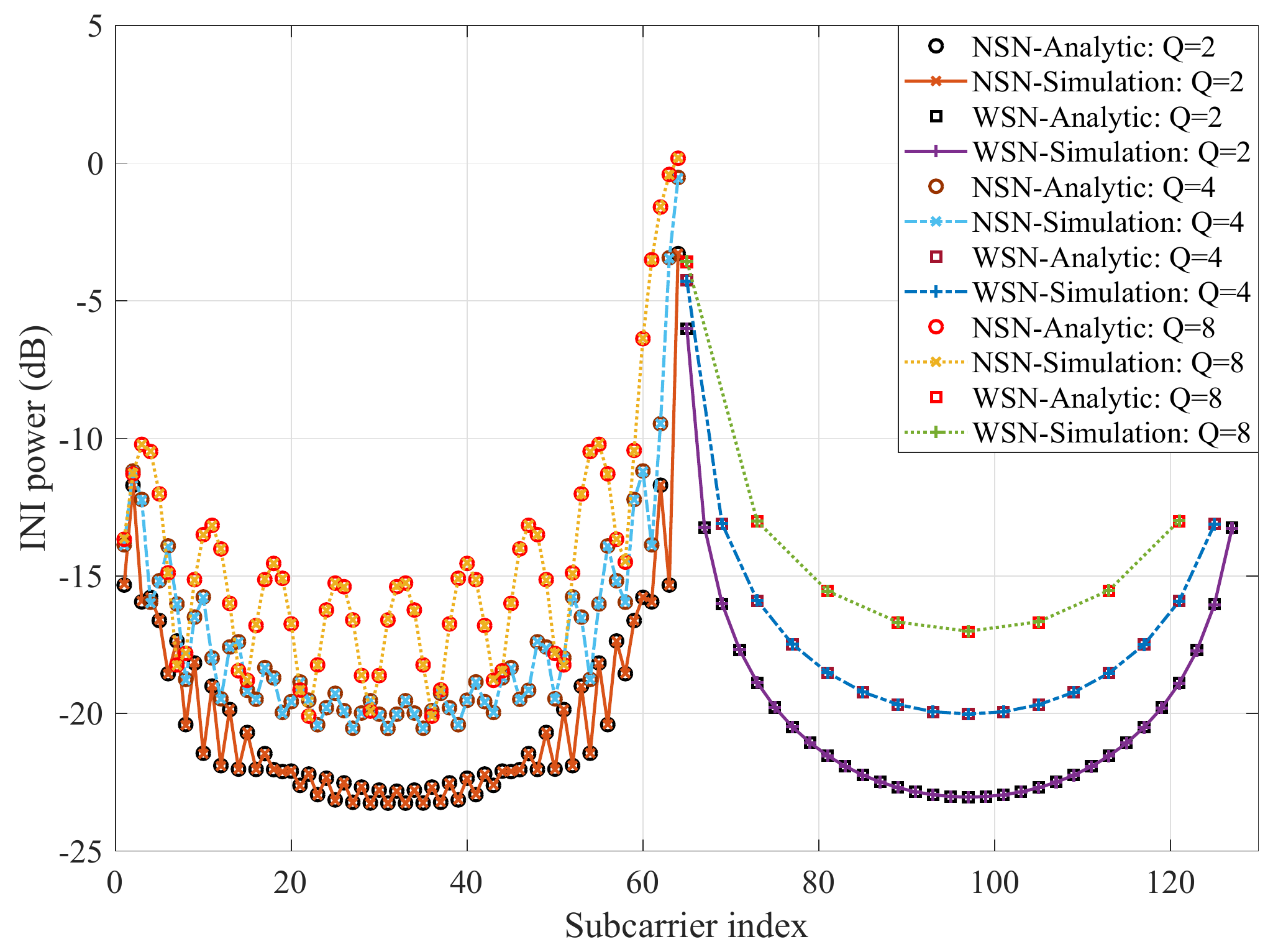} \label{fig:VNAsetup}
	}	
	\subfigure[Common CP]{	
		\includegraphics[width=0.965\columnwidth]{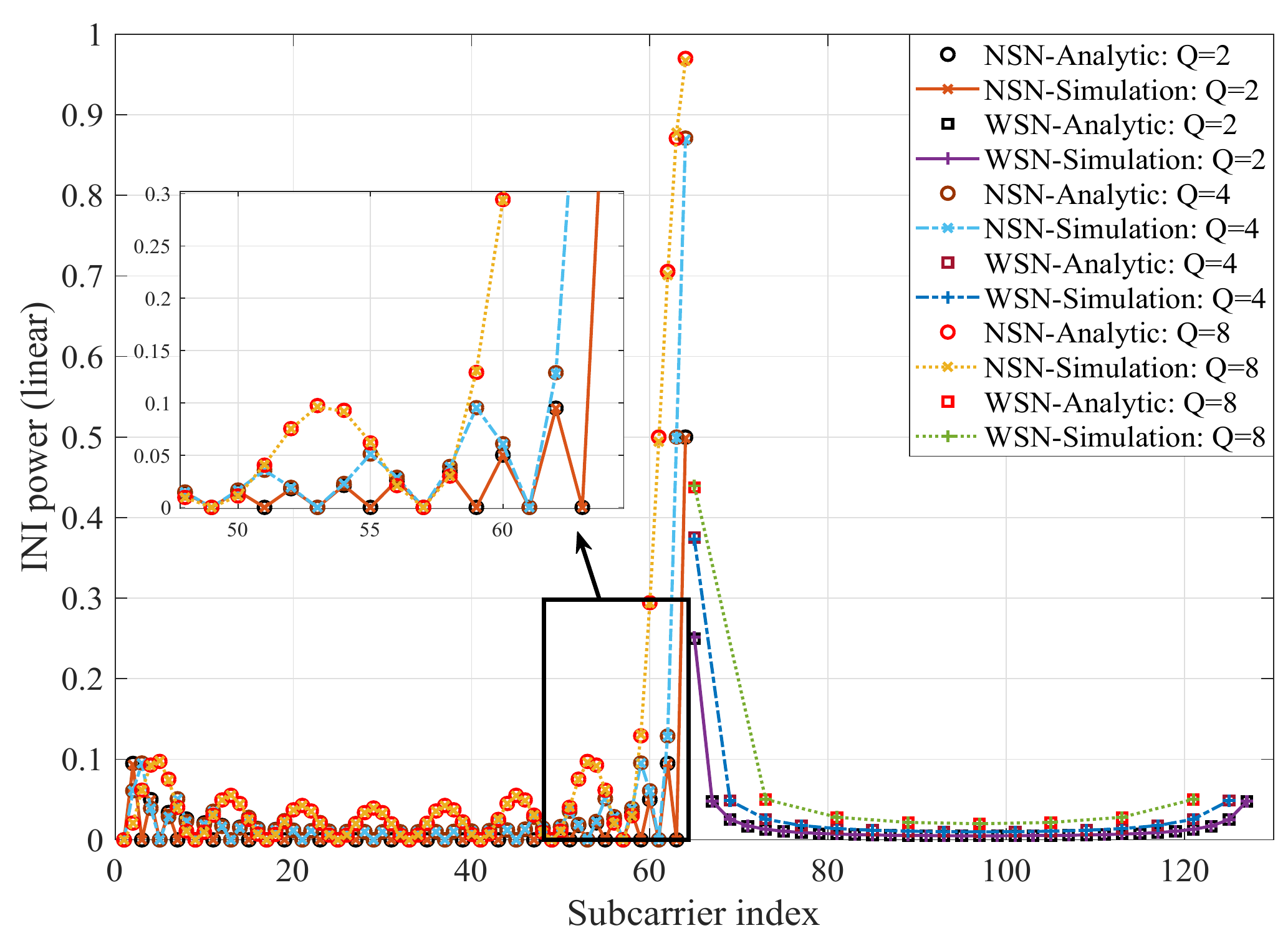} \label{fig:DopplerMeasurement}
	}
	\caption{INI as a function of $Q$}
	\label{fig:DifferentQ}
\end{figure*}

\subsection{INI comparison for individual and common CP}
Illustrated in Fig. \ref{fig:IndividualCommonCP} is the INI power experienced by each subcarrier of NSN and WSN for both, individual and common CP configurations. As one could expect, the edge subcarriers of both numerologies are experiencing relatively higher INI compared to the middle subcarriers. This is due to the higher sidelobes of the adjacent numerology, whose effect abates as one moves away from the edges. Again, as observed from the analytical expressions \eqref{eq:INIEq25} and  \eqref{eq:INIEq32}, the INI experienced by WSN is exactly the same for both CP configurations, and that agrees well with the simulation results shown in Fig. \ref{fig:IndividualCommonCP}. The INI of NSN exhibits an interesting oscillatory behavior from one subcarrier to another. This behavior is due to the discontinuities created at the boundaries of the $Q$-concatenating symbols of WSN. That is why, as it can be observed from the next figure (i.e., Fig. \ref{fig:DifferentQ}), this oscillation repeats itself after every $Q$ subcarriers. The behavior is even more interesting when common CP configuration is used whereby one out of $Q$ subcarriers suffers zero INI. The same observation is reported in \cite{zhang2017subband} for multi-numerology system utilizing universal filtered multicarrier waveform. This peculiar INI distribution exhibited by NSN can be of great advantage in some aspects. For instance, with common CP configuration, pilot subcarriers for channel estimation or subcarries of the reliability-sensitive application can be distributed to those locations with zero INI. Another critical thing to be noted from Fig. \ref{fig:IndividualCommonCP} is that, for NSN subcarriers with non zero INI with common CP case, experiences higher INI compared to when individual CP configuration is used. Therefore, the fact that common CP configuration has an advantage of having number of INI-free subcarriers comes at a cost of increased INI on the other subcarriers. 

\subsection{INI as a function of subcarrier spacing ratio ($Q$)}
Let us investigate the effect of the ratio of different subcarrier spacing for the two numerologies. Fig. \ref{fig:DifferentQ} shows INI distribution on each subcarrier of NSN and WSN for different value of $Q$. Here, from the definition of $Q$ (i.e., $Q = \Delta f_2/\Delta f_1$), we fix $\Delta f_1$ at 15 kHz and change $\Delta f_2$ from 30 kHz, 60 kHz and 120 kHz. The INI each numerology receives seems to increase with the increase in $Q$. This is quite expected since the higher the $\Delta f_2$ the larger the sidelobes of the WSN that impart higher INI on NSN subcarriers. On the other hand, increasing $\Delta f_2$ widens the spectrum of the WSN and thus collects more interference from NSN.

\begin{figure}[t]
	\centering
	\includegraphics[width=0.96\columnwidth]{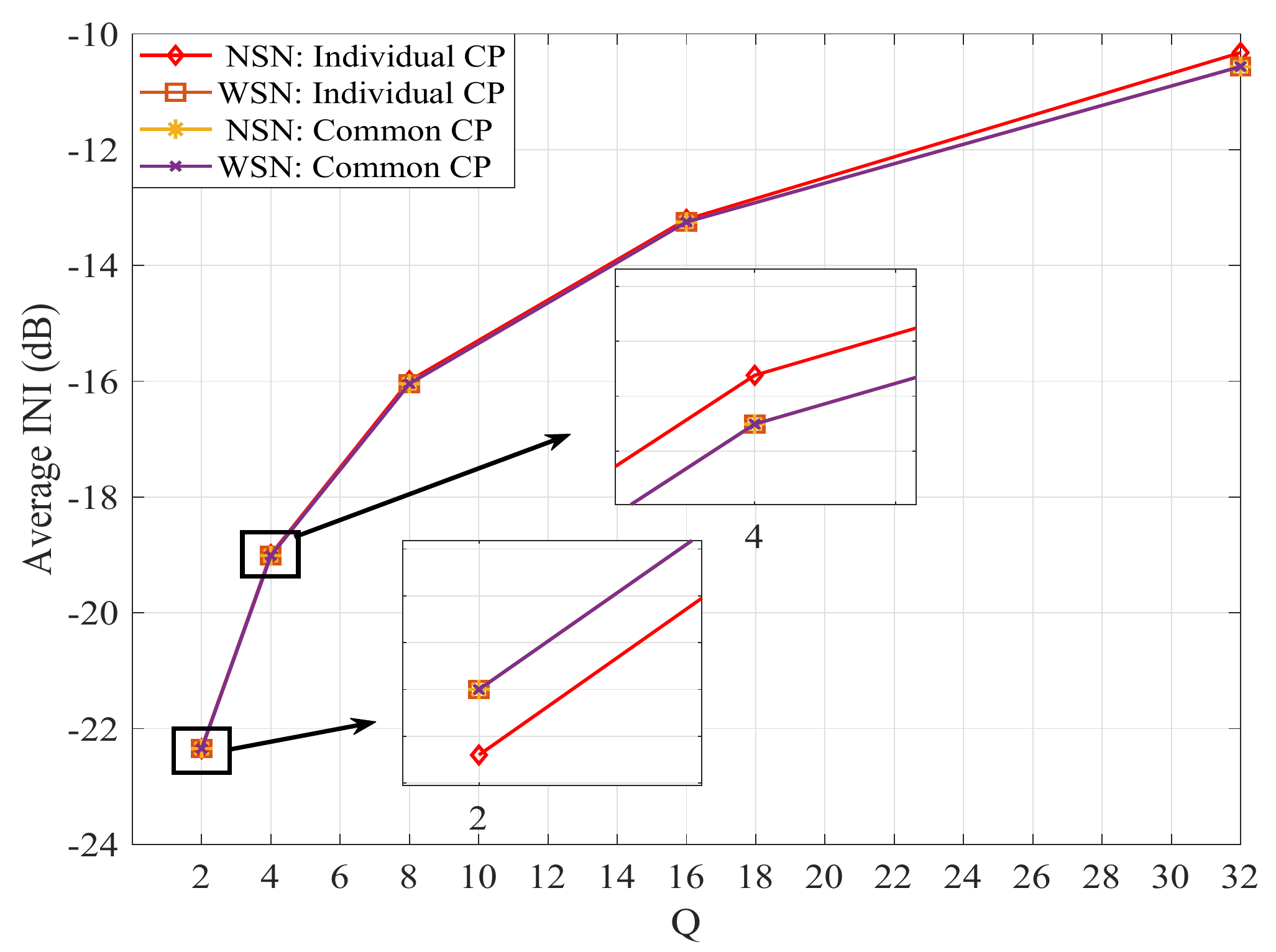}
	\caption{Average INI experienced by each numerology with individual and common CP for different values of Q}
	\label{fig:AverageINIQ}
\end{figure}
\begin{figure}[t]
	\centering
	\includegraphics[width=0.95\columnwidth]{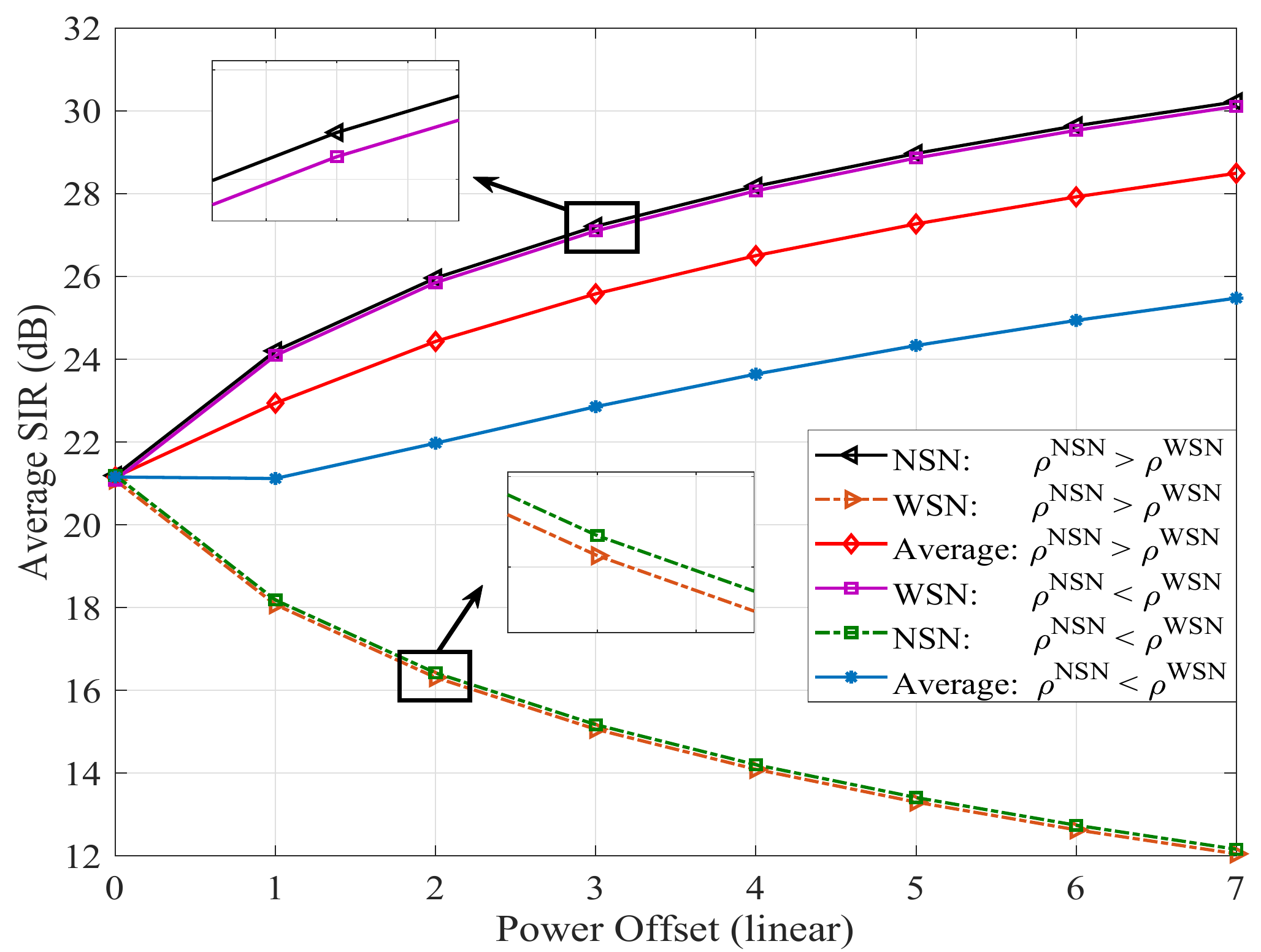}
	\caption{Average SIR of the multi-numerology system as a function of the power offset}
	\label{fig:PowerOffset}
\end{figure}

As mentioned above, INI oscillation in NSN becomes more pronounced with increasing $Q$. For the common CP configuration, one $Q$-th of all NSN subcarriers are INI free. Although the INI distribution among subcarriers of NSN and WSN with common and individual CP configurations is quite diverse, it is observed that, for a given $Q$, if  $\rho^{NSN} = \rho^{WSN}$ (i.e., no power offset between the two numerologies), average INI of NSN and WSN with common CP, and WSN with individual CP are exactly the same. The average INI of NSN with individual CP is slightly different from the rest, as shown in Fig. \ref{fig:AverageINIQ}. We count that this deviation is due to the dependency of the INI of NSN with individual CP on the $CP_R$ (see \eqref{eq:INIEq17}). That is, the CP portion of each of the $Q$-concatenated WSN symbols contributes to the INI experienced by NSN. In Fig. \ref{fig:AverageINIQ} we can see that the average INI of NSN with individual CP is smaller than the rest when $Q$ = 2, and larger when $Q \geq$ 4. The deviation is observed to be even more substantial with higher $CP_R$ and higher $Q$.

\subsection{Effect of power offset on INI} 
Depending on the channel conditions and application, users can have diverse power requirement, leading to power offsets among them within the system. Power offset among the users utilizing the same numerology does not impose any threat on the performance of the system since the orthogonality is maintained. However, in the multi-numerology system, where INI exists, power offset between users utilizing different numerologies highly affects the SIR performance of the system \cite{kihero2018inter}. Fig. \ref{fig:PowerOffset} presents SIR performance of the multi-numerology system utilizing individual CP configuration as a function of power offset. The investigation is done for $Q$ = 2, and two cases, $\rho^{NSN} > \rho^{WSN}$ and $\rho^{WSN} > \rho^{NSN}$, are considered. For each case, average SIR performance of each numerology as well as average SIR of the whole system are observed. It is clear from Fig. \ref{fig:PowerOffset} that power offset favors the numerology with higher power and degrades performance of the other numerology with relatively lower power. This is obvious since numerology with higher power has stronger sidelobes whose interference to the adjacent numerology is more severe. We note from Fig. \ref{fig:PowerOffset} that, the amount by which SIR of NSN is improved and that of WSN is degraded when $\rho^{NSN} > \rho^{WSN}$ is exactly the same as the amount by which SIR of WSN is improved and that of NSN is degraded when $\rho^{WSN} > \rho^{NSN}$. However, for a given power offset, performance of NSN is slightly larger than that of WSN. This is only because of using $Q$ = 2. As we have discussed earlier, when $Q$ = 2, average INI of NSN is slightly less compared to that of WSN (see Fig. \ref{fig:AverageINIQ}). It is also observed from Fig. \ref{fig:PowerOffset} that increasing power of one numerology relative to the other improves the overall average SIR of the system. However, this improvement is found to depend on which numerology has higher power. The average system SIR is higher for the case with $\rho^{NSN} > \rho^{WSN}$ than $\rho^{WSN} > \rho^{NSN}$. The reason is that, for a given bandwidth, NSN has more subcarriers that WSN, making its effect on the average system SIR greater than that of WSN. Therefore, it is worth mentioning that, the observed trend of the average system SIR is subjected to change when the two multiplexed numerologies do not share the available bandwidth equally (i.e., $\eta_1 \neq \eta_2$). We therefore understand that, the overall SIR performance of the system is simultaneously affected by both, power offset and bandwidth sharing factor, $\eta$.

From the above discussion it is evident that, even though power offset among users of different numerologies may improve the overall system performance, it however causes severe degradation in the performance of the numerology with relatively lower power. In order to favor both numerologies, it is more plausible to maintain the power offset between them as low as possible. 

\section{Conclusion} \label{sec:conclusion}  
In this paper, a thorough investigation of INI for CP-OFDM waveform is done through simulation and analytical derivations. The most critical factors affecting the amount of INI suffered in the system are analyzed and some guidelines of minimizing their effects are highlighted. For instance an intelligent scheduling of the multiplexed numerologies can be used to minimize INI in the first place without adopting spectrally inefficient techniques like guard band and windowing. Placing numerologies with the minimum subcarrier spacing ratio adjacent to each other, minimizing the power offset between adjacent numerologies as well as making a proper choice of mixed numerologies symbol alignment techniques (i.e., individual or common CP) in time domain are found to be some of the inherent ways of controlling INI in the system.     

\appendices
\section{Simplification of \eqref{eq:INIEq16}}\label{sec:AppendixA}
Let us consider the first term of \eqref{eq:INIEq16}
\begin{equation}\label{eq:INIEqAppA1}
\begin{split}
 &\sum_{r=N_{CP}}^{M+M_{CP}-1}y_{w_0}(r)\mathrm{e}^{-j2\pi rv/N}\\&
= \sum_{r=0}^{M+M_{CP}-N_{CP}-1}y_{w_0}(r+N_{CP})\mathrm{e}^{-\frac{j2\pi}{N}(r+N_{CP})v}\\&
\stackrel{(a)}{=}\frac{1}{\sqrt{M}}\sum_{r=0}^{M+M_{CP}-N_{CP}-1}\sum_{k=0}^{\eta_2N-1}X_{w_0}(\eta_1M + k/Q)\\& \quad \quad \quad \quad \quad \quad \quad 
\cdot \mathrm{e}^{\frac{j2\pi}{N}(r+N_{CP})(k+\eta_1N)}\mathrm{e}^{-\frac{j2\pi}{N}v(r+N_{CP})}\\&
= \frac{1}{\sqrt{M}}\sum_{k=0}^{\eta_2N-1}X_{w_0}(\eta_1M + k/Q)\mathrm{e}^{\frac{j2\pi}{N} N_{CP}(k-v+\eta_1N)}\\& \quad \quad \quad \quad \quad \quad \quad
\cdot \sum_{r=0}^{M+M_{CP}-N_{CP}-1}\mathrm{e}^{-\frac{j2\pi}{N}r(k-v+\eta_1N)}\\&
\stackrel{(b)}{=}\frac{1}{\sqrt{M}}\sum_{k=0}^{\eta_2N-1}X_{w_0}(\eta_1M + k/Q) \mathrm{e}^{\frac{j2\pi}{N} N_{CP}(k-v+\eta_1N)}\\& \quad \quad \quad \quad \quad \quad \quad  
\cdot \frac{1-\mathrm{e}^{\frac{j2\pi}{N}(M+M_{CP}-N_{CP})(k-v+\eta_1N)}}{1-\mathrm{e}^{\frac{j2\pi}{N} (k-v+\eta_1N)}}\\&
\stackrel{(c)}{=}\frac{1}{\sqrt{M}}\sum_{k=0}^{\eta_2 N-1}X_{w_0}(\eta_1M + k/Q) \mathrm{e}^{\frac{j2\pi}{N} N_{CP}(k-v+\eta_1N)}\\& \quad \quad \quad \quad \quad \quad \quad \quad 
\cdot \mathrm{e}^{\frac{j\pi}{N} (M + M_{CP}-N_{CP}-1)(k-v+\eta_1N)}\\& \quad \quad \quad
\cdot \frac{sin \bigg[\frac{\pi}{N}(M+M_{CP}-N_{CP})(k-v)\bigg]}{sin \bigg[\frac{\pi}{N}(k-v+\eta_1N)\bigg]},
\end{split}
\end{equation}
where the equality $(a)$ is obtained by evaluating the term $y_{w_0}(r+N_{CP})$ by using \eqref{eq:INIEq5}. The equalities $(b)$ and $(c)$ follow after using the formula for the sum of a finite geometric series on the summation over $r$-indices, and the Euler's formula, respectively.

By using \eqref{eq:INIEq5} and following the same steps as above, the second term of \eqref{eq:INIEq16} can be expressed as
\begin{equation}\label{eq:INIEqAppA2}
\begin{split}
&\sum_{r=0}^{M+M_{CP}-1}\sum_{q=1}^{Q-1}y_{w_q}(r-q(M+M_{CP}))\mathrm{e}^{\frac{-j2\pi}{N}v (r-q(M+M_{CP}))}\\&
= \sum_{q=1}^{Q-1}\frac{1}{\sqrt{M}}\sum_{k=0}^{\eta_2 N-1}X_{w_q}(\eta_1M + k/Q)\\& \quad \quad \quad
\cdot \mathrm{e}^{-\frac{j2\pi}{N} (M+M_{CP})(k-v+\eta_1N)}
\mathrm{e}^{\frac{j\pi}{N} (M + M_{CP}-1)(k-v+\eta_1N)}\\& \quad \quad \quad
\cdot \frac{sin \bigg[\frac{\pi}{N}(M+M_{CP})(k-v)\bigg]}{sin \bigg[\frac{\pi}{N}(k-v+\eta_1N)\bigg]}.
\end{split}
\end{equation}

\section{Simplification of \eqref{eq:INIEq24}}\label{sec:AppendixB}
Substituting \eqref{eq:INIEq4} into \eqref{eq:INIEq24}, we have
\begin{equation}\label{eq:INIEqAppA3}
\begin{split}
&Y_{INI}(\eta_1M + p/Q)\\&
= \frac{1}{\sqrt{M\cdot N}}\sum_{r=0}^{M-1}\sum_{k=0}^{\eta_1N-1}X_{nr}(k)\mathrm{e}^{\frac{j2\pi}{N} k(r+M_{CP}-q(M+M_{CP}))}\\& \quad \quad \quad \quad \quad \quad \quad \quad \quad \quad \quad \quad \quad
\cdot \mathrm{e}^{-\frac{j2\pi}{M}(r+M_{CP})(\eta_1M + p/Q)}\\&
= \frac{1}{\sqrt{M\cdot N}}\sum_{k=0}^{\eta_1N-1}X_{nr}(k)\mathrm{e}^{-\frac{j2\pi}{N} k(M_{CP}-q(M+M_{CP}))}\\& \quad \quad \quad \quad \quad
\cdot \mathrm{e}^{-\frac{j2\pi}{M}(M_{CP}(\eta_1M + p/Q))} \sum_{r=0}^{M-1}\mathrm{e}^{-\frac{j2\pi}{N}r(k-p-\eta_1N)}\\&
\stackrel{(d)}{=}\frac{1}{\sqrt{M\cdot N}}\sum_{k=0}^{\eta_1N-1}X_{nr}(k)\mathrm{e}^{-\frac{j2\pi}{N} k(M_{CP}-q(M+M_{CP}))}\\& \quad \quad \quad \quad \quad
\cdot \mathrm{e}^{-\frac{j2\pi}{M}(M_{CP}(\eta_1M + p/Q))}\cdot \frac{1-\mathrm{e}^{\frac{j2\pi}{N}M(k-p-\eta_1N)}}{1-\mathrm{e}^{\frac{j2\pi}{N} (k-p-\eta_1N)}}\\&
\stackrel{(e)}{=}\frac{1}{\sqrt{M\cdot N}}\sum_{k=0}^{\eta_1N-1}X_{nr}(k)\mathrm{e}^{-\frac{j2\pi}{N} k(M_{CP}-q(M+M_{CP}))}\\& \quad \quad \quad \quad \quad
\cdot \mathrm{e}^{-\frac{j2\pi}{M}(M_{CP}(\eta_1M + p/Q))}\mathrm{e}^{-\frac{j2\pi}{N} (M-1)(k-p-\eta_1N)}\\&
\quad \quad \quad \quad \quad
\cdot \frac{sin \bigg[\frac{\pi}{Q}(k-p)\bigg]}{sin \bigg[\frac{\pi}{N}(k-p-\eta_1N)\bigg]},
\end{split}
\end{equation}
where the equalities $(d)$ and $(e)$ again follow after employing the formula for the sum of a finite geometric series on the summation over $r$-indices, and Euler's formula, respectively. 


\begin{IEEEbiography}[{\includegraphics[width=1in,height=1.25in,clip,keepaspectratio]{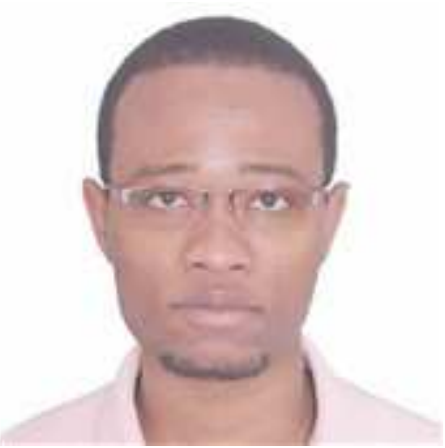}}]{Abuu B. Kihero}
	received his B.S. degree in Electronics engineering from Gebze Technical University, Kocaeli, Turkey in 2015, and M.S. degree in in Electrical, Electronics and Cyber systems from Istanbul Medipol University, Istanbul, Turkey in 2018. He is currently working as a research assistant at Communication, Signal Processing and Networking Center (CoSiNC) while pursuing his Ph.D degree in wireless communication technologies at Istanbul Medipol University. 
\end{IEEEbiography}
\begin{IEEEbiography}[{\includegraphics[width=1in,height=1.25in,clip,keepaspectratio]{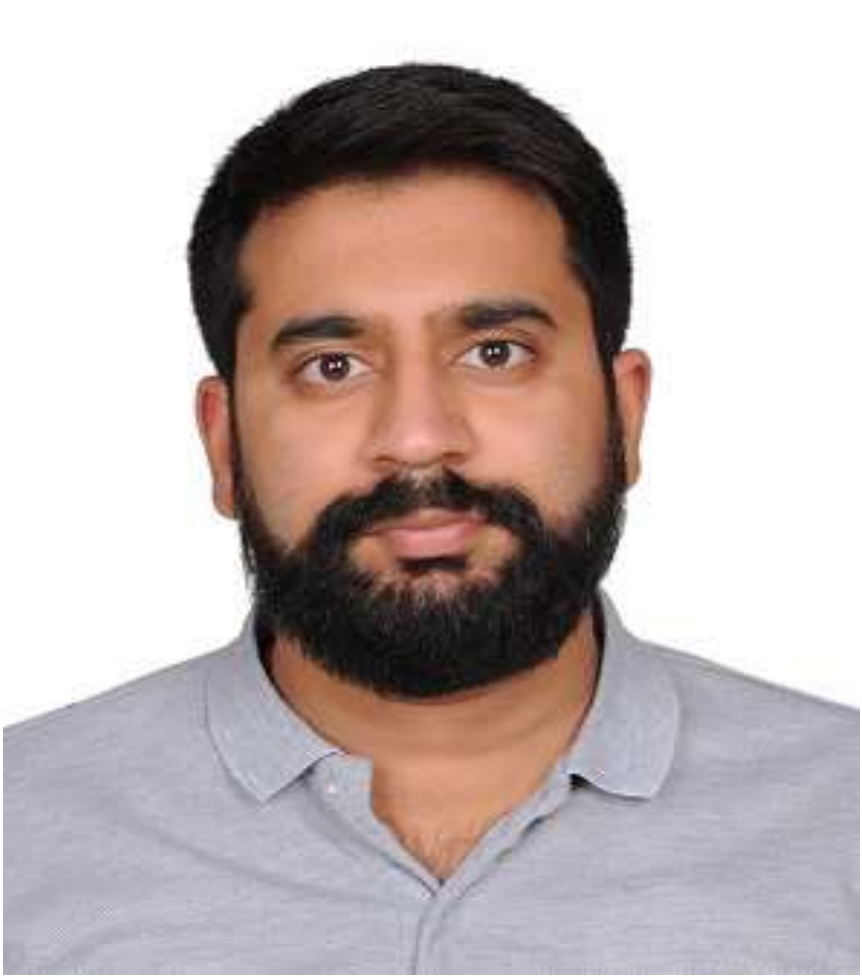}}]{Muhammad Sohaib J. Solaija} (S’16) received his B.E and M.Sc degrees in electrical engineering from National University of Science and Technology, Islamabad, Pakistan in 2014 and 2017, respectively. Currently he is pursuing his Ph.D. degree as a member of the Communications, Signal Processing, and Networking Center (CoSiNC) at Istanbul Medipol University, Turkey. His research focuses on interference modelling and coordinated multipoint implementation for 5G and beyond wireless systems.
\end{IEEEbiography}
\begin{IEEEbiography}[{\includegraphics[width=1in,height=1.25in,clip,keepaspectratio]{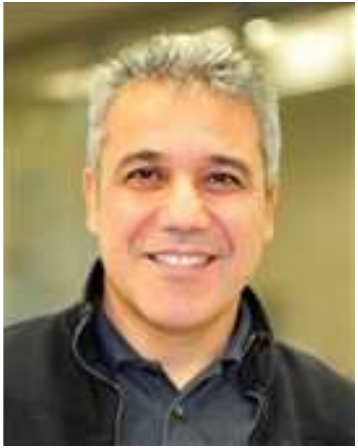}}]{H{\"u}seyin Arslan}(S’95-M’98-SM’04-F’15) has received his BS degree from Middle East Technical University (METU), Ankara, Turkey in 1992; MS and PhD. degrees in 1994 and 1998 from Southern Methodist University (SMU), Dallas, TX, USA. From January 1998 to August 2002, he was with the research group of Ericsson Inc., NC, USA, where he was involved with several project related to 2G and 3G wireless communication systems. Since August 2002, he has been with the Electrical Engineering Dept. of University of South Florida, Tampa, FL, USA. Also, he has been the dean of the College of Engineering and Natural Sciences of Istanbul Medipol University since 2014. In addition, he has worked as part time consultant for various companies and institutions including Anritsu Company (Morgan Hill, CA, USA), The Scientific and Technological Research Council of Turkey (TUBITAK). His current research interests are on physical layer security, mmWave communications, small cells, multi-carrier wireless technologies, co-existence issues on heterogeneous networks, aeronautical (high altitude platform) communications and in vivo channel modeling, and system design. 
\end{IEEEbiography}
	
\end{document}